\documentclass[aps,prb,twocolumn,superscriptaddress,floatfix]{revtex4}
 
\advance\textheight by 1.truecm

\newcommand{\bk}{{\bm k}}

\usepackage{xspace}
\newcommand{\K}{K\xspace}
\newcommand{\Kp}{K$^\prime$\xspace}
\newcommand{\M}{M\xspace}

\usepackage{graphicx,graphics}
\usepackage{dcolumn}
\usepackage{amsmath,amssymb,amsfonts}
\usepackage{latexsym,verbatim}
\usepackage{bm}
\usepackage{color}
\usepackage{ulem}
\usepackage[percent]{overpic}
\usepackage[breaklinks=true,colorlinks,citecolor=blue,linkcolor=blue,urlcolor=blue]{hyperref}
\usepackage{multirow}

\begin{document}
\title{Spin-resolved optical conductivity of two-dimensional group-VIB transition-metal dichalcogenides}

\author{Marco Gibertini}
\email{marco.gibertini@epfl.ch}
\affiliation{Theory and Simulation of Materials (THEOS) and National Center for Computational Design and Discovery of Novel Materials (MARVEL), \'Ecole Polytechnique F\'ed\'erale
de Lausanne, CH-1015 Lausanne, Switzerland}
\author{Francesco M.D. Pellegrino}
\email{francesco.pellegrino@sns.it}
\affiliation{NEST, Scuola Normale Superiore and Istituto Nanoscienze-CNR, I-56126 Pisa, Italy}
\author{Nicola Marzari}
\affiliation{Theory and Simulation of Materials (THEOS) and National Center for Computational Design and Discovery of Novel Materials (MARVEL), \'Ecole Polytechnique F\'ed\'erale
de Lausanne, CH-1015 Lausanne, Switzerland}
\author{Marco Polini}
\affiliation{NEST, Istituto Nanoscienze-CNR and Scuola Normale Superiore, I-56126 Pisa, Italy}
\affiliation{Istituto Italiano di Tecnologia, Graphene Labs, Via Morego 30, I-16163 Genova, Italy}
\begin{abstract}
We present an ab-initio study of the spin-resolved optical conductivity of two-dimensional (2D) group-VIB transition-metal dichalcogenides (TMDs). We carry out fully-relativistic density-functional-theory calculations combined with maximally localized Wannier functions to obtain band manifolds at extremely high resolutions and focus on the photo-response of 2D TMDs to circularly-polarized light in a wide frequency range. We present extensive numerical results for monolayer TMDs involving molybdenum and tungsten combined with sulphur and selenium. Our numerical approach allows us to locate with a high degree of accuracy the positions of the points in the Brillouin zone that are responsible for van Hove singularities in the optical response. Surprisingly, some of the saddle points do not occur exactly along high-symmetry directions in the Brillouin zone, although they happen to be in their close proximity.
\end{abstract}

\maketitle

\section{Introduction}
\label{sec:introduction}

Graphene-related materials~\cite{novoselov_ps_2012,bonaccorso_matertoday_2012} (GRMs) are at the center of a great deal of experimental and theoretical interest~\cite{ferrari_nanoscale_2014,novoselov_nature_2012,koppens_naturenano_2014,grigorenko_naturephoton_2012} 
because of their potential impact in technological applications ranging from electronics to photonics, opto-electronics, and plasmonics.

Among GRMs, two-dimensional (2D) transition metal dichalcogenides (TMDs)~\cite{novoselov_pnas_2005} have emerged as a class of promising materials~\cite{wang_naturenano_2012,chhowalla_naturechem_2013}. 
Contrary to 2D semimetals such as single-layer graphene, 2D TMDs are semiconductors~\cite{cappelluti_prb_2013,rostami_prb_2013,kormanyos_prb_2013,carvalho_prb_2013,roldan_2dmat_2014,yuan_prb_2014} 
with a direct gap in the visible frequency range, which makes them ideal candidates for applications requiring strong light-matter interactions~\cite{mak_prl_2010,splendiani_nanolett_2010,britnell_science_2013}.

Monolayers of group-VIB TMDs are compounds of the type ${\rm MX}_2$ where ${\rm M}$ is a transition metal such as ${\rm Mo}$ and ${\rm W}$ and ${\rm X}$ is a chalcogen such as ${\rm S}$ and ${\rm Se}$. Structurally, 2D group-VIB TMDs are composed of an atomic trilayer in a ${\rm X}$-${\rm M}$-${\rm X}$ configuration where ${\rm M}$ displays a trigonal prismatic coordination; the whole crystal is characterized by hexagonal symmetry. Conduction and valence band edges are at the corners \K and \Kp of the first Brillouin zone (${\rm 1BZ}$).
At variance with graphene, these 2D materials are non-centrosymmetric and they are characterized by strong spin-orbit coupling (SOC), which is mostly due to the presence of heavy metals~\cite{kormanyos_prb_2013,roldan_2dmat_2014}.

The lack of inversion symmetry leads to a low-energy valley optical selection rule, with inter-band transitions around the 
\K and \Kp valleys being coupled to left- and right-handed circularly polarized light~\cite{xiao_prl_2012}, respectively.
SOC together with the lack of inversion symmetry induces strong coupling between spin and
valley of electrons. These features can lead to the possibility of technological 
applications based on the control of spin and valley degrees-of-freedom~\cite{mak_naturenano_2012,cao_naturecommun_2012,zeng_naturenano_2012,jones_naturenano_2013,ross_naturenano_2014}.

Using density functional theory (DFT) calculations, Carvalho {\it et al.}~\cite{carvalho_prb_2013} were the first to evaluate the joint density-of-states (JDOS) of 2D TMDs. 
They found logarithmic singularities---i.e.~van Hove singularities (VHSs)---in the JDOS as a function of energy and attributed them to ``band nesting'' (see below). 
Evidence of VHSs has recently emerged in photoluminescence excitation spectroscopy studies~\cite{kozawa_naturecommun_2014} of 2D TMDs.

In this Article we investigate the spin-resolved optical conductivity of selected 2D TMDs taking into account SOC. We use fully-relativistic ab-initio DFT calculations 
{\it combined} with the use of maximally localized Wannier functions. This approach allows us to study with great accuracy the 
critical points in the band structure of 2D TMDs and to locate saddle points, which are responsible for VHSs in the optical spectra.

This Article is organized as following. Technical details on the first-principles calculations are summarized in Sect.~\ref{sec:technicaldetails}. 
In Sect.~\ref{sec:bands} we present our main results for the fully-relativistic band structure together with a discussion of the impact of SOC. In Sect.~\ref{sec:opt} we collect our main results for the optical conductivity as a function of photon energy and for its dependence on the light polarization. Sect.~\ref{sec:VHS} is devoted to band nesting and VHSs. A brief summary and our main conclusions are reported in Sect.~\ref{sec:end}.

\section{Technical details on the first-principles calculations}
\label{sec:technicaldetails}
First-principles simulations have been performed at the level of DFT, as implemented in the PWscf code of the Quantum ESPRESSO distribution~\cite{giannozzi_jpcm_2009}. The exchange-correlation energy functional has been approximated either using the local-density approximation~\cite{lda_prb_1981} (LDA) or the generalized gradient approximation (GGA), as introduced by Perdew, Burke and Ernzerhof~\cite{pbe_prl_1996} (PBE). 
Although Kohn-Sham electronic band structures are not meant to represent true quasiparticle excitations (for which GW many-body perturbation theory~\cite{Giuliani_and_Vignale} needs to be applied on top of DFT) the topology of the band manifolds is typically well reproduced, with the notable exception of the band gap.

Electron-ion interactions are treated using fully-relativistic optimized norm-conserving Vanderbilt pseudopotentials~\cite{hamann_prb_2013} that include semi-core states for transition metals. An energy cutoff of ${\rm 65}~{\rm Ryd}$ is used to expand wavefunctions into plane waves, together with a gamma-centered $12\times12\times1$ Monkhorst-Pack grid to sample the 1BZ. In order to deal with 2D systems using a plane-wave basis set, a $20~{\rm \AA}$ layer of vacuum has been considered to minimize interaction between periodic replicas. We adopted the Broyden-Fletcher-Goldfarb-Shanno method to 
 relax forces on atoms below $3~{\rm meV}/{\rm \AA}$ for different lattice constants and then obtain equilibrium geometries using the corresponding equation of state.
Maximally localized Wannier functions~\cite{marzari_rmp_2012} were constructed for the top fourteen valence bands (including spin) and bottom eight conduction bands using the Wannier90 code~\cite{mostofi_cpc_2008}. Projections over $d$-orbitals of the transition metal and $p$-orbitals of the chalcogens have been employed as starting points for the localization procedure. In order to compute accurately and inexpensively the optical properties of 2D TMDs, we exploited Wannier-function techniques~\cite{marzari_rmp_2012} to interpolate the band structure on a very fine grid in momentum space. Brillouin zone integrations were then carried out using the tetrahedron method~\cite{blochl_prb_1994}.

\section{Band structure and the role of spin-orbit coupling}
\label{sec:bands}

In this Section we present the electronic band structure of group-VIB monolayer 
TMDs of the type ${\rm MX}_2$ with ${\rm M} = {\rm Mo}, {\rm W}$ and 
${\rm X} = {\rm S}, {\rm Se}$.

\begin{figure*}[t]
\includegraphics[width=\textwidth]{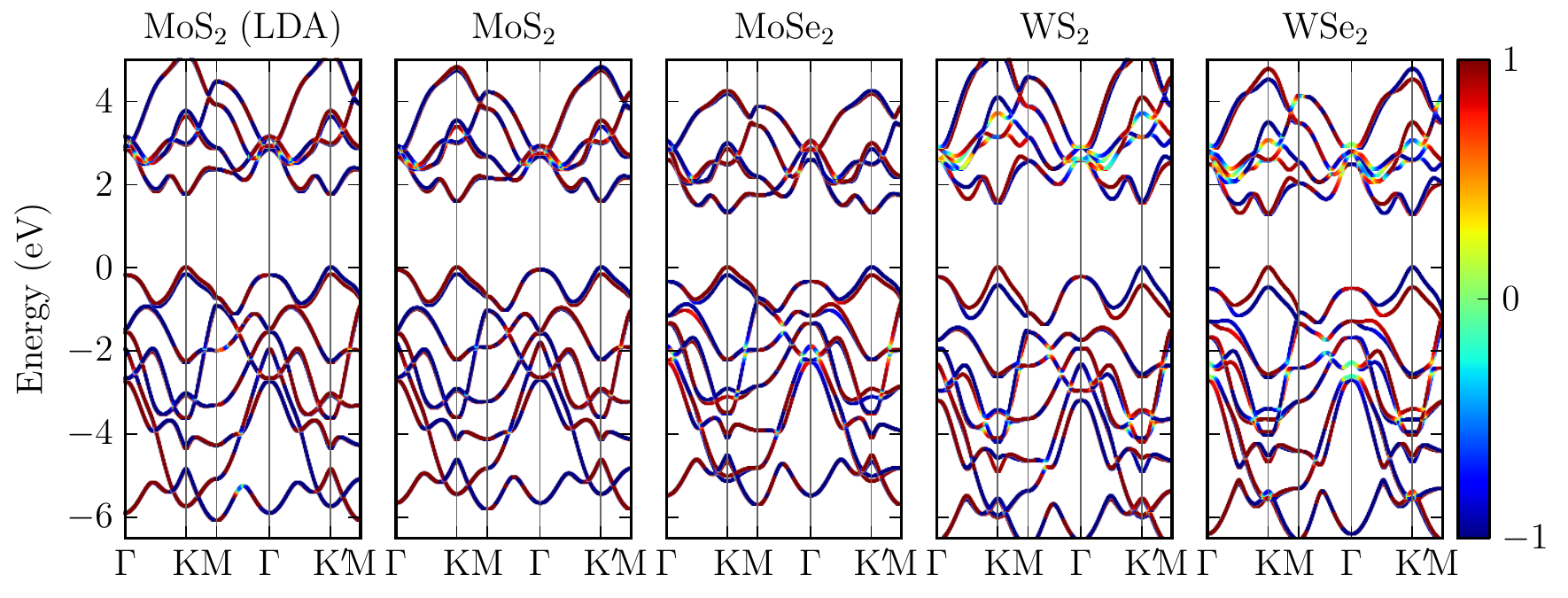}
\caption{(Color online) Wannier-interpolated DFT results for the band structure of four selected 2D TMDs. Bands are plotted along the high-symmetry path $\Gamma$-\K-\M-$\Gamma$-\Kp.
The color of the solid lines indicates the value of the spin projection along the ${\hat {\bm z}}$ direction. Red (blue) indicates spin pointing up (down). 
In the case of ${\rm MoS}_2$, we plot Wannier-interpolated DFT band structures obtained both at the LDA (leftmost panel) and GGA levels. 
For the other 2D crystals we only present GGA results.\label{fig:W}}
\end{figure*}
Fig.~\ref{fig:W} shows the fully relativistic band structure of 2D TMDs obtained using the Wannier interpolation approach~\cite{marzari_rmp_2012}, together with the expectation value of the projection of the spin operator ${\hat S}_z$ along the ${\hat {\bm z}}$ direction. The zero of energy has been set at the top of the valence bands for undoped 2D TMDs, so that electron bands corresponding to negative energy values are fully filled (valence), while electron bands corresponding to positive energy values are empty (conduction). 

For all TMDs considered here, we plot the Wannier-interpolated band structure obtained from GGA calculations, while in the case of ${\rm MoS}_2$ also LDA results are reported (leftmost panel). The equilibrium lattice constants $a$ adopted in the underlying DFT calculations are shown in Table~\ref{tab:comp},  together with the equilibrium distances $d$ between chalcogen atoms in the unit cell. By comparing the band structures of ${\rm MoS}_2$ obtained by using these approximate exchange-correlation functionals we find only minor quantitative (but not qualitative) differences. For instance, we note that the GGA direct energy gap ($1.59$~eV) is smaller than the LDA one ($1.78$~eV). This is true also for the others TMDs, as from Table~\ref{tab:comp}. 
We remark that such differences simply arise from the different crystal structures (i.e.~values of the parameters $a$ and $d$) predicted using LDA or GGA. Indeed, we have verified that both LDA and GGA give practically identical results for the band manifolds when identical crystal structures are used in the calculations.
It is very well known~\cite{Giuliani_and_Vignale}, however, that both approximations understimate the true energy gap. 
Recent calculations, which include electron-electron interaction corrections~\cite{shi_prb_2013,louie_prl_2013,mollina_prb_2013} by means of the GW approximation~\cite{Giuliani_and_Vignale}, predict an energy gap that is roughly $1~{\rm eV}$ larger than the energy gap obtained by using e.g.~GGA. The GW gap compares well with experimental data in the case of ${\rm MoSe}_2$~\cite{ugeda_nmat_2014}. 
This correction to the gap needs to be kept in mind when one compares 
our GGA data for the optical conductivity in Fig.~\ref{fig:opt} below with experimental data.

In addition to the energy gap $E_{\rm g}$ at the 1BZ corners, in Table~\ref{tab:comp} we compare LDA and GGA results for the spin-orbit energy splitting at \K for all TMDs considered here. While for Mo-based TMDs we find almost perfect agreement, slightly larger discrepancies can be seen when W is involved, with LDA predicting smaller values for the valence band splitting $\Delta_{\rm SO}^{\rm v}$ and larger values for the conduction band splitting $\Delta_{\rm SO}^{\rm c}$ with respect to GGA.  Our results are in agreement with what has been previously reported in the literature at the LDA~\cite{kormanyos_prx_2014} and GGA~\cite{kosmider_prb_2013,liu_prb_2013} levels.

\begin{table}
\caption{Comparison between LDA and GGA-PBE results for several quantities: equilibrium lattice constant $a$ (in \AA);  distance between chalcogen atoms $d$ (in \AA);  spin-orbit energy splitting at \K (in meV) for the top valence band, $\Delta_{\rm SO}^{\rm v}$, and bottom conduction band, $\Delta_{\rm SO}^{\rm c}$; energy gap at \K (in eV). Sign conventions for spin-orbit energy splittings are in agreement with Refs.~\onlinecite{kosmider_prb_2013,liu_prb_2013}. \label{tab:comp}}
\begin{tabular}{l  c c c c c c c c c c c}
\toprule
			& \multicolumn{5}{c }{LDA} & &\multicolumn{5}{c}{GGA} \\
			\cline{2-6} \cline{8-12}
			& $a$ &  $d$ & $\Delta^{\rm v}_{\rm SO}$ & $\Delta^{\rm c}_{\rm SO}$ & $E_{\rm g}$  & &
			$a$ & $d$ & $\Delta^{\rm v}_{\rm SO}$ & $\Delta^{\rm c}_{\rm SO}$ & $E_{\rm g}$ \\
\colrule			
MoS$_2$		& 3.12 & $3.11$	& $146$ & $-3$  & $1.78$ & & 3.19 & $3.13$	& $149$ & $-3$  & $1.59$ \\
MoSe$_2$	& 3.25 & $3.32$	& $186$ & $-23$ & $1.52$ & & 3.32 & $3.34$	& $186$ & $-21$ & $1.34$ \\
WS$_2$		& 3.12 & $3.12$	& $414$ & $\phantom{-}37$  & $1.74$ & & 3.18 & $3.14$	& $427$ & $\phantom{-}29$  & $1.56$ \\
WSe$_2$		& 3.24 & $3.33$	& $439$ & $\phantom{-}47$  & $1.47$ & & 3.32 & $3.36$	& $464$ & $\phantom{-}37$  & $1.27$ \\		 
\botrule
\end{tabular}
\end{table}

We note that in Fig.~\ref{fig:W}, independently of the specific approximation for the exchange-correlation energy functional, spin-flip effects due to SOC are negligible~\cite{kormanyos_prb_2013,roldan_2dmat_2014} in the explored 2D TMDs.
In other words, ${\hat S}_z$ is an {\it approximate quantum number} despite the fact that the projection ${\hat S}_z$ of the spin operator along the ${\hat {\bm z}}$ direction does {\it not} commute with the Hamiltonian when SOC is included. We now elaborate on this issue a bit more in depth. To this aim, let us first consider some properties of the eigenstates of the Hamiltonian in the absence of SOC. In this case, electron bands have a twofold spin degeneracy and the expectation value of ${\hat S}_z$ is a good quantum number. 
A generic eigenstate of the Hamiltonian without SOC can be written as 
$|\lambda, s, \bk \rangle$, where $\lambda$ is a band index, $s$ is the eigenvalue of ${\hat S}_z$, and $\bk$ is the crystal momentum. 
Owing to the symmetry under horizontal mirror-plane inversion, $\hat\sigma_h$, electron bands can be classified~\cite{cappelluti_prb_2013} as either even (E) or odd (O). Even (odd) bands 
satisfy $\hat\sigma_h |\lambda_{\rm E}, s, \bk \rangle = +|\lambda_{\rm E}, s, \bk \rangle$  
($ \hat\sigma_h |\lambda_{\rm O}, s, \bk \rangle = -|\lambda_{\rm O}, s, \bk \rangle$). 

In the atomic approximation, SOC has the following form~\cite{kormanyos_prb_2013}
\begin{equation}\label{eq:soc}
 \hat{\cal H}_{\rm SO} = \frac{\hbar^2}{4 m_{\rm e}^2 c^2} \frac{1}{r} \frac{d V(r)}{d r} \hat{\bm L} \cdot \hat{\bm S},
\end{equation}
where $m_{\rm e}$ is the electron mass, $V(r)$ is the spherical atomic potential, $\hat{\bm L}$ is the angular momentum operator, and $\hat{\bm S}$ is a vector of spin-$1/2$ Pauli matrices
(with eigenvalues $\pm 1$).
Under the symmetry operation $\hat\sigma_h$, the SOC Hamiltonian remains invariant, so that $\hat\sigma_h$ is still a symmetry. 
By writing  
\begin{equation}\label{eq:sh}
\hat{\bm L} \cdot \hat{\bm S}  = \hat L_+\hat S_- + \hat L_-\hat S_+ + \hat L_z \hat S_z
\end{equation}
with $\hat L_\pm=\hat L_x\pm i \hat L_y$ and $\hat S_\pm=(\hat S_x\pm i \hat S_y)/2$, 
one easily obtains that 
\begin{equation}\label{eq:ldots}
\langle  \lambda, s, \bk |\hat{\bm L} \cdot \hat{\bm S} |\lambda^\prime ,s^\prime, \bk \rangle=\langle \lambda, s, \bk |\hat L_z \hat S_z |\lambda^\prime, s^\prime,  \bk \rangle~,
\end{equation}
if the bands $\lambda$ and $\lambda^\prime$ have the same spin ($s=s^\prime$). Eq.~(\ref{eq:ldots}) is different from zero only when the bands have the same parity with respect to $\hat\sigma_h$. Indeed, $\hat L_z$ commutes with $\hat\sigma_h$ and the states $|\lambda^\prime, s,  \bk \rangle$ are eigenstates of $\hat S_z$, so that 
\begin{align}
\langle \lambda, s, \bk |\hat L_z \hat S_z |\lambda^\prime, s,  \bk \rangle &= s \langle \lambda, s, \bk | \hat L_z |\lambda^\prime, s,  \bk \rangle\nonumber\\
&= s \langle \lambda, s, \bk |\hat\sigma_h^{-1} \hat L_z \hat\sigma_h|\lambda^\prime, s,  \bk \rangle\nonumber\\
&= \pm s \langle \lambda, s, \bk | \hat L_z |\lambda^\prime, s,  \bk \rangle \nonumber\\
&= \pm \langle \lambda, s, \bk | \hat L_z \hat S_z |\lambda^\prime, s,  \bk \rangle~,
\end{align}
where the upper (lower) sign holds if the states have the same (opposite) parity under the action of $\hat\sigma_{h}$.
Analogously, if the states have opposite spin ($s^\prime={\bar s}$), we find:

\begin{equation}
\langle \lambda, s, \bk |\hat{\bm L} \cdot \hat{\bm S} | \lambda^\prime, {\bar s},  \bk \rangle=\langle \lambda, s, \bk |\hat L_+\hat S_- + \hat L_-\hat S_+ |\lambda^\prime, {\bar s},  \bk \rangle~,
\end{equation}
which is non-vanishing only when the states have different parity since $\hat L_{\pm}$ anti-commutes with $\hat\sigma_h$. In the spirit of first-order perturbation theory, we therefore conclude that SOC may induce a spin mixing only if the corresponding energy scale is comparable with the energy separation between two bands with opposite parity. Since SOC energy scales are typically small, we have that the spin-flip terms can usually be neglected~\cite{kormanyos_prb_2013,feng_prb_2012,cappelluti_prb_2013,liu_prb_2013} and the SOC Hamiltonian (\ref{eq:soc}) can therefore be approximated as a mere Zeeman-type spin splitting
\begin{equation}\label{eq:zeeman}
\hat{\cal H}_{\rm SO} \approx \frac{\hbar^2}{4 m_{\rm e}^2 c^2} \frac{1}{r} \frac{d V(r)}{d r} \hat L_z \hat S_z~.
\end{equation}
 A more sophisticated analysis beyond first-order perturbation theory~\cite{kosmider_prb_2013,kormanyos_prx_2014} shows that, although SOC can still be treated as a Zeeman-type splitting $\hat{\cal H}_{\rm SO} \propto \hat S_z$ as in Eq.~\eqref{eq:zeeman}, the proportionality factor might have significant contributions from coupling to virtual states. This is particularly relevant for the lowest conduction band where these additional contributions are responsible for the different sign of the spin-orbit splitting $\Delta_{\rm SO}^{\rm c}$ for MoX$_2$ and WX$_2$ crystals\cite{kosmider_prb_2013,kormanyos_prx_2014} in Table~\ref{tab:comp}.
The approximate conservation of $\hat S_z$ according to Eq.~\eqref{eq:zeeman} is confirmed in Fig.~\ref{fig:W}, where the expectation value of ${\hat S}_z$ is almost everywhere well defined and equal to $\langle {\hat S}_z \rangle= \pm 1$. The only exceptions occur close to energy crossings between bands with opposite parity, where the expectation value of the spin-flip term is no longer negligible with respect to the energy separation between the bands and gives rise to an avoided crossing together with a rotation of the spin expectation from $\langle {\hat S}_z \rangle= + 1$ to $\langle {\hat S}_z \rangle= -1$ (or viceversa). Thus, to a large extent, each band index $n$, obtained by diagonalizing the full Hamiltonian including the SOC term in Eq.~(\ref{eq:soc}), can be associated with a doublet $\lambda, s$, where $\lambda$ (or any other Greek letter such as $\mu, \nu$) denotes a band index in the absence of SOC (as above) and $s = \uparrow,\downarrow$ is the eigenvalue of ${\hat S}_z$.

Before concluding this Section, we would like to highlight that SOC does not lift the twofold degeneracy of the bands when the crystal momentum $\bk$ spans the high-symmetry line $\Gamma$-\M. Indeed, the small point group~\cite{Dresslhaus} of each crystal momentum $\bk$ belonging to the high-symmetry line $\Gamma$-\M is $C_{2v}$: this group contains the identity, the mirror reflection $\hat{\sigma}_h$, the rotation of an angle $\pi$ around an axis along the $\Gamma$-\M line, and the mirror reflection $\hat{\sigma}_v$ with respect to the vertical plane containing the line $\Gamma$-\M. By including spin, the only physically relevant representation of the double group of $C_{2v}$ that distinguishes the identity from rotations of an angle $2\pi$ is two-dimensional. As a consequence, even though spin-mixing is allowed close to energy crossings (see above), the bands remain always twofold degenerate along the $\Gamma$-\M line.

\section{Numerical results: spin-resolved optical conductivity}
\label{sec:opt}

In this Section we study the optical response~\cite{yates_prb_2007,ebert_rep_1996} of monolayer TMDs to normally incident  monochromatic light by calculating the real part of the optical conductivity.

\subsection{Formal aspects}

The optical conductivity tensor $\hat{\bm \sigma}(\omega)$ has to be invariant with respect to any crystal symmetry~\cite{yates_prb_2007,ebert_rep_1996}. 
For monolayer TMDs we find that $\hat{\bm \sigma}(\omega)$ needs to be of the form
\begin{equation}\label{eq:tensor}
\hat{\bm \sigma}(\omega)=\sigma_{\parallel}(\omega) \hat{\openone} + \sigma_{\perp}(\omega) \hat{\tau}_y~,
\end{equation}
where $\hat{\openone}$ is the identity matrix and $\hat{\tau}_y$ is a Pauli matrix. 
The entries of these matrices refer to the Cartesian components of the optical conductivity tensor.
In Eq.~(\ref{eq:tensor}), $\sigma_{\parallel} $ and $\sigma_{\perp}$ are complex functions of the frequency of the external electric field. Specifically, the optical absorbance is related to the real part of the optical conductivity tensor $\hat{\bm \sigma}$. For instance, in the presence of linearly polarized light, impinging perpendicularly to the monolayer plane, 
the optical absorbance is proportional to 
$\Re e(\sigma_{\parallel})$, and is independent of the direction of polarization. 
In the presence of circularly polarized light, the off-diagonal component $\sigma_{\bot}$ modifies the optical absorbance with respect to the case of linear polarization. 
For circularly polarized light, the real part of the optical conductivity is defined as
\begin{equation}\label{eq:re}
\Re e[\sigma_{\pm}] = \Re e [ \sigma_{\parallel}] \pm \Re e [\sigma_{\perp}]~, 
\end{equation}
where $+$ ($-$) refers to left (right) circular polarization.

\begin{figure}[t]
\includegraphics{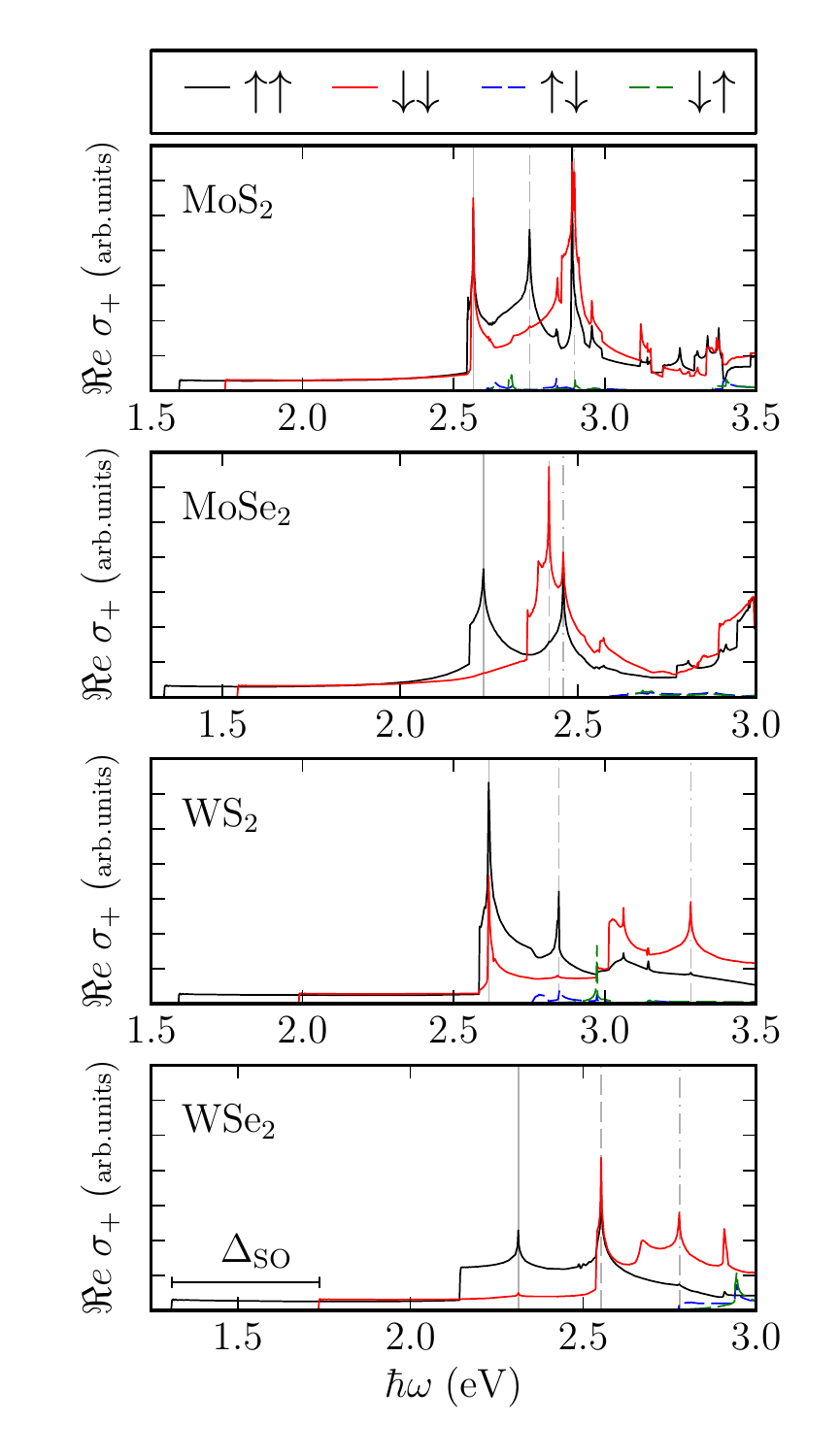}
\caption{(Color online) Optical absorbance (in arbitrary units) as a function of photon energy $\hbar \omega$ (in ${\rm eV}$) 
for the same 2D TMDs as in Fig.~\ref{fig:W}.
Solid lines refer to the diagonal spin components, $\Re e~[\sigma^{\uparrow\uparrow}_{+}(\omega)]$ (black) and $\Re e~[\sigma^{\downarrow\downarrow}_{+}(\omega)]$ (red).  
Dashed lines refer to the spin-flip components, $\Re e~[\sigma^{\uparrow\downarrow}_{+}(\omega)]$ (blue) and $\Re e~[\sigma^{\downarrow\uparrow}_{+}(\omega)]$ (green).
Vertical lines denote the three energies $E_{1}$ (solid), $E_{2}$ (dashed), and $E_{3}$ (dash-dotted), at which low-energy VHSs occur. All the results in this figure have been obtained at the GGA-PBE level. \label{fig:opt}}
\end{figure}

The real part of the optical conductivity can be evaluated from 
the Kubo formula~\cite{Giuliani_and_Vignale} in the dipole approximation, i.e.
\begin{widetext}
\begin{equation} \label{eq:spm}
\Re e[ \sigma_{ \pm}(\omega)]=\frac{\pi e^2}{2 \omega} \sum^{\rm occ}_{n}  \sum^{\rm empty}_{m} \int_{\rm 1BZ} \frac{ d^2 \bk}{(2 \pi)^2}
 \delta (\hbar \omega-\varepsilon_{m, \bk }+\varepsilon_{n, \bk }) |\langle m, \bk|{\hat v}_\pm | n, \bk\rangle|^2, 
\end{equation}
\end{widetext}
where the 2D integral is carried over the 1BZ, ${\hat v}_\pm={\hat v}_x \pm i {\hat v}_y$, and ${\hat {\bm v}} = (\hat{v}_x,\hat{v}_y)$ is the velocity operator, which is defined by 
${\hat {\bm v}} = - i[{\hat {\bm r}}, {\hat {\cal H}}]/\hbar$. Here ${\hat {\cal H}}$ is the full Hamiltonian, including SOC, and $|n, \bk\rangle$ are the corresponding Bloch eigenstates. Using Wannier functions, the matrix elements of the velocity operator together with the eigenenergies $\varepsilon_{n,\bk}$ can be interpolated over a very fine $\bk$-space grid in a very efficient and inexpensive way~\cite{marzari_rmp_2012,yates_prb_2007}. 
Below we present calculations at zero temperature and in the intrinsic (undoped) limit.

By exploiting the correspondence $n  \leftrightarrow (\nu,s)$ discussed above in Sect.~\ref{sec:bands}, we can partition the optical conductivity in Eq.~(\ref{eq:spm}) in the sum of four spin-resolved contributions:
\begin{equation}\label{eq:spin}
  \sigma_{ \pm}(\omega) = \sum_{s= \uparrow, \downarrow}  \sum_{s^\prime=\uparrow,\downarrow}  \sigma^{ss^\prime}_{ \pm}(\omega)~.
\end{equation}
The meaning of $\sigma^{ss^\prime}_{ \pm}(\omega)$ is the following. Let us, for example, consider the case $s = \uparrow$ and $s^\prime = \downarrow$. In this case, the quantity $\sigma^{\uparrow\downarrow}_{\pm}(\omega)$ is computed including in Eq.~\eqref{eq:spm} only inter-band processes from an empty band with spin-$\downarrow$ to an occupied band with spin-$\uparrow$. Since $\hat S_z$ is only approximately conserved, the spin character of a given state $|n,\bk\rangle$ is assigned according to the expectation value $\langle n\bk|\hat S_z|n\bk\rangle$, with spin-$\uparrow$ corresponding to  $\langle n\bk|\hat S_z|n\bk\rangle > 0$ and spin-$\downarrow$ to $\langle m\bk|{\hat S}_z|m\bk\rangle <0$. 

Such a spin-resolved analysis, allows us to study the spin polarization of the photo-current in response to circularly polarized light. In addition, exploiting time reversal symmetry (TRS), one obtains the following relation between the spin-resolved components of the optical conductivity:
\begin{equation}\label{eq:right}
\Re e[\sigma^{s s^\prime}_{\pm} (\omega)] =  \Re e[\sigma^{\bar{s} \bar{s}^\prime}_{\mp}(\omega)]~,
\end{equation}
where $\bar{s}=\downarrow$ ($\bar{s}=\uparrow$) if $s=\uparrow$ ($s=\downarrow$). For this reason, in what follows we present results for the case of the optical response to left-handed light only. However, before concluding this Section, we briefly discuss the properties of the optical response to linearly-polarized light. 

Taking into account the relation between the optical responses to left- and right-handed circularly-polarized light 
in Eq.~(\ref{eq:right}), it is clear that the spin-polarization of the photo-response can be reversed by reversing the light polarization.
Finally, we remind the reader that the optical response $\sigma^{s s^\prime}_{\parallel}$ to linearly-polarized light can be expressed as
\begin{equation}\label{eq:linearpolarization}
\Re e[\sigma^{s s^\prime}_{\parallel}] = \frac{1}{2}(\Re e[\sigma^{s s^\prime}_{+}] + \Re e[ \sigma^{s s^\prime}_{-}])~.
\end{equation}
By exploiting TRS as from Eq.~(\ref{eq:right}), one finds 
\begin{equation}\label{linearpolarization_2}
\Re e[\sigma^{s s^\prime}_{\parallel}] = \frac{1}{2}(\Re e[\sigma^{s s^\prime}_{+}] + \Re e[ \sigma^{\bar{s} \bar{s}^\prime}_{+}])~.
\end{equation}
Eq.~(\ref{linearpolarization_2}) implies that the optical response to linearly-polarized light does not depend on the projection 
of the spin operator ${\hat S}_z$ along the ${\hat {\bm z}}$ direction. In other words, the photo-current generated by linearly-polarized light 
does not carry spin polarization.

\subsection{Numerical results and discussion}

Fig.~\ref{fig:opt} illustrates the spin-resolved contributions to $\Re e[\sigma_+(\omega)]$ (in arbitrary units) in response to left-handed light, as functions of the excitation energy $\hbar \omega$ (in ${\rm eV}$). Results in this figure have been obtained by using the GGA-PBE exchange and correlation energy functional. 
We have checked that these results do not change qualitatively upon changing approximation for the exchange-correlation potential (e.g.~by doing LDA).

We immediately see that the spin-flip contributions to $\Re e[\sigma_+(\omega)]$, i.e. $\Re e[\sigma_+^{\uparrow \downarrow}(\omega)]$ and $\Re e[\sigma_+^{\downarrow \uparrow}(\omega)]$, are vanishingly small at low energies and negligible with respect to the spin-diagonal contributions, i.e.~$\Re e[\sigma_+^{\uparrow \uparrow}(\omega)]$ and $\Re e[\sigma_+^{\downarrow \downarrow}(\omega)]$, at high energies. This is a consequence of the very weak spin-flip effect induced by SOC, as discussed in Sect.~\ref{sec:bands}.

From now on, we will therefore focus our attention on the spin-diagonal contributions only.
At a generic value of the photon energy $\hbar\omega$, we clearly see that the spin-diagonal contributions 
$\Re e[\sigma_+^{\uparrow \uparrow}(\omega)]$ and $\Re e[\sigma_+^{\downarrow \downarrow}(\omega)]$ are very different.
This immediately implies that the photo-response to circularly-polarized light is substantially spin-polarized. 
In particular, there is an energy window $\Delta_{\rm SO} = \Delta_{\rm SO}^{\rm v} -  \Delta_{\rm SO}^{\rm c}$ at low energies  in which the resultant spin polarization is $100\%$ ($\Delta_{\rm SO} = 152~{\rm meV}$ for ${\rm MoS}_{2}$, $\Delta_{\rm SO} = 206~{\rm meV}$ for ${\rm MoSe}_2$, $\Delta_{\rm SO} = 398~{\rm meV}$ 
for ${\rm WS}_{2}$, $\Delta_{\rm SO} = 426~{\rm meV}$ for ${\rm WSe}_{2}$). This is precisely the same energy range in which an optical spin-valley selection rule exists~\cite{xiao_prl_2012, li_prb_2012,yao_prb_2008,rostami_prb_2014}. This particular result can also by derived by using a low-energy effective model, e.g.~the massive Dirac model, which can be obtained by expanding the electronic structure around the two principal valleys \K and ${\rm K}^{\prime}$, in close proximity to the conduction- and valence-band edges.
The quantity $\Delta_{{\rm SO}}$ is a measure of the strength of SOC. In agreement with earlier literature~\cite{roldan_2dmat_2014}, 
we find that the 2D TMD with the largest $\Delta_{\rm SO}$ is ${\rm WSe}_{2}$. This is because ${\rm WSe}_2$ is composed by the heaviest metal ($Z_{\rm W}=74$, $Z_{\rm Mo} = 42$) and {\it also} by the heaviest chalcogen ($Z_{\rm Se} = 34$, $Z_{\rm S}=16$).

All 2D TMDs studied in this work display a step-like increase of the optical absorbance for sufficiently large energies (e.g. $\hbar \omega \gtrsim  2.54~{\rm eV}$ for ${\rm MoS}_2$). This sudden jump stems from a point in the 1BZ along the $\Gamma$-\K direction where a secondary minimum appears in conduction band, as shown in Fig.~\ref{fig:W}. (We remind the reader that the absolute minimum occurs at the \K point.) This disconnected pocket where electrons promoted by light from valence band can end up is responsible for the step-like increase of the optical absorbance. 

With increasing photon energy, one sees that the optical conductivity presents a series of logarithmic divergences, i.e.~VHSs, at which a large increase of the optical absorbance occurs. In every panel of Fig.~\ref{fig:opt} we have denoted by vertical lines the locations of the three most relevant VHSs with energy $E_1$ (solid), $E_2$ (dashed), and $E_3$ (dash-dotted). 
As we will discuss below, these VHSs correspond to transitions involving the top valence bands and the bottom conduction band for each spin.

Both 2D TMDs containing sulphur, i.e.~${\rm MoS}_2$ and ${\rm WS}_2$, display the lowest VHS at an energy which is independent of the spin polarization, 
$E_{1} = 2.565~{\rm eV}$ and $E_{1} = 2.616~{\rm eV}$, respectively. On the contrary, the second VHS in ${\rm MoS}_2$ and ${\rm WS}_2$ occurs for spin-$\uparrow$ (spin-$\downarrow$) polarization for left-handed (right-handed) light, at energies $E_{2} = 2.751~{\rm eV}$ and $E_{2} = 2.848~{\rm eV}$, respectively. At these energies, the photo-excited electrons in ${\rm MoS}_2$ (${\rm WS}_2$) possess a degree of spin-$\uparrow$ polarization ${\cal P}_\uparrow \simeq 71\%$ (${\cal P}_\uparrow \simeq 81\%$), with
\begin{equation}\label{eq:degree-of-spin-polarization}
{\cal P}_s \equiv \frac{\Re e[\sigma_+^{ss}]}{\Re e[ \sigma_+]}~.
\end{equation}
Finally, for  MoS$_2$ (WS$_2$) the VHS at $E_3 = 2.900~{\rm eV}$ ($3.284 ~{\rm eV}$) occurs for spin-$\downarrow$ electrons with a partial polarization ${\cal P}_\downarrow \simeq 68\%$ (${\cal P}_\downarrow \simeq 75\%$). In the case of MoS$_2$, this VHS is very close in energy to another spin-degenerate VHS associated with transitions close to the $\Gamma$-\M line involving  for  each spin the third conduction band in this region of the 1BZ.

We finally discuss the optical response of TMDs containing selenium, i.e.~${\rm MoSe}_2$ and ${\rm WSe}_2$. In this case even the first VHS yields a strongly spin-polarized photo-current. For these compounds we have $E_{1} = 2.234~{\rm eV}$ and $E_{1} = 2.312~{\rm eV}$, respectively. At these energies, the photo-excited electrons in ${\rm MoSe}_2$ (${\rm WSe}_2$) possess a degree of spin-$\uparrow$ polarization ${\cal P}_\uparrow \simeq 84\%$ (${\cal P}_\uparrow \simeq 82\%$). 

\begin{figure}[t]
\includegraphics{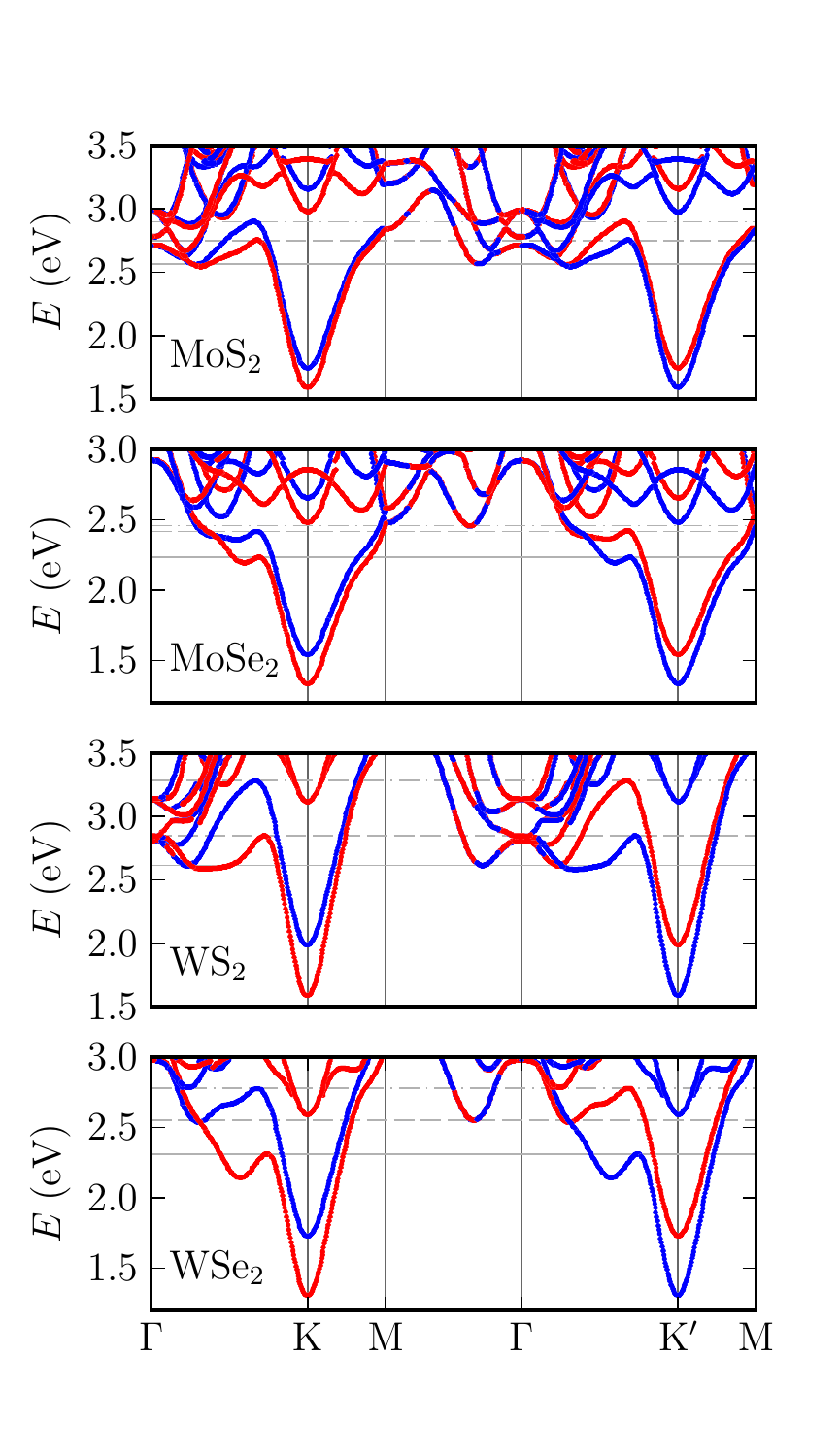}
\caption{(Color online) Wannier-interpolated GGA-PBE band energy differences $\varepsilon_{\mu, s,\bk}-\varepsilon_{\nu, s,\bk}$ 
for 2D TMDs along the high-symmetry path $\Gamma$-\K-\M-$\Gamma$-\Kp-\M.
Red (blue) lines refer to the difference $\varepsilon_{\mu, \uparrow, \bk}-\varepsilon_{\nu, \uparrow, \bk}$ ($\varepsilon_{\mu, \downarrow, \bk}-\varepsilon_{\nu, \downarrow, \bk}$). Horizontal lines denote the three energy values $E_{1}$ (solid), $E_{2}$ (dashed), and $E_{3}$ (dash-dotted), at which VHSs occur in the optical conductivity plots shown in Fig.~\ref{fig:opt}.\label{fig:band_diff}}
\end{figure}
\section{Band nesting and van Hove singularities}
\label{sec:VHS}

In this Section we discuss the origin of VHSs in the optical conductivity. As discussed in Ref.~\onlinecite{carvalho_prb_2013}, these are due to the phenomenon of ``band nesting''. 

``Band nesting'' refers to the presence of regions in the 1BZ where the occupied band $\varepsilon_{n,\bk}$ can be obtained from an empty band 
$\varepsilon_{m,\bk}$ by means of a rigid vertical shift in an energy-momentum band diagram. 
These regions occur in proximity of a point in ${\bm k}$-space where the {\it 2D gradient of the energy difference} $\varepsilon_{m,\bk}-\varepsilon_{n,\bk}$ vanishes, 
i.e.~$\nabla_{\bk} (\varepsilon_{m,\bk}-\varepsilon_{n,\bk})= 0$. The point in ${\bm k}$-space where this condition is met can either be 
a saddle point or an extremum (a minimum or maximum). In a 2D system band nesting involving a saddle point induces a logarithmic singularity in the optical conductivity~\cite{bassani_pastori}. On the other hand, if band nesting involves an extremum, the corresponding optical conductivity exhibits a step-like behavior.

We therefore need to study with extreme care the energy differences $\varepsilon_{m, \bk}-\varepsilon_{n, \bk}$, where $n$ ($m$) is the label of an occupied (empty) band. Fig.~\ref{fig:band_diff} illustrates such differences for the four 2D TMDs of interest in this work, plotted along the high-symmetry path $\Gamma$-\K-\M-$\Gamma$-\Kp-\M. In this figure, red (blue) lines refer to the energy difference $\varepsilon_{\mu, \uparrow,\bk}-\varepsilon_{\nu, \uparrow, \bk}$ 
 ($\varepsilon_{\mu, \downarrow, \bk}- \varepsilon_{\nu, \downarrow, \bk}$). Here, the meaning of the Greek labels $\mu, \nu$ is identical to that explained earlier in Sect.~\ref{sec:bands} and in the context of Fig.~\ref{fig:W}. Differences between band energies with opposite values of the spin projection along the ${\hat {\bm z}}$ direction have been discarded since, as we have discussed above, spin-flip processes play a marginal role.  TRS ($\varepsilon_{\lambda, \uparrow,\bk}=\varepsilon_{\lambda, \downarrow,-\bk}$) is evident 
in that the energy differences along the $\Gamma$-\K-\M and $\Gamma$-\Kp-\M directions 
are identical {\it modulo} a reversal of the projection of the spin along the ${\hat {\bm z}}$ direction (or color flip ${\rm red} \leftrightarrow {\rm blue}$ in the figure). 
The absence of a spin splitting along the \M-$\Gamma$ direction, which we have discussed earlier in Sect.~\ref{sec:bands}, is also apparent.

Horizontal lines in Fig.~\ref{fig:band_diff} correspond to the three energies $E_{1}$, $E_{2}$, and $E_{3}$ 
at which VHSs occur in the optical conductivity---see Fig.~\ref{fig:opt}. These energies lie very close to local extrema in the energy difference between the top valence band and the bottom conduction band for each spin. This means that at least one component of the gradient $\nabla_{\bk} (\varepsilon_{m,\bk}-\varepsilon_{n,\bk})$ is vanishing and that we can focus on such bands to better understand the nature of the VHSs.

\begin{figure}[t]
\includegraphics{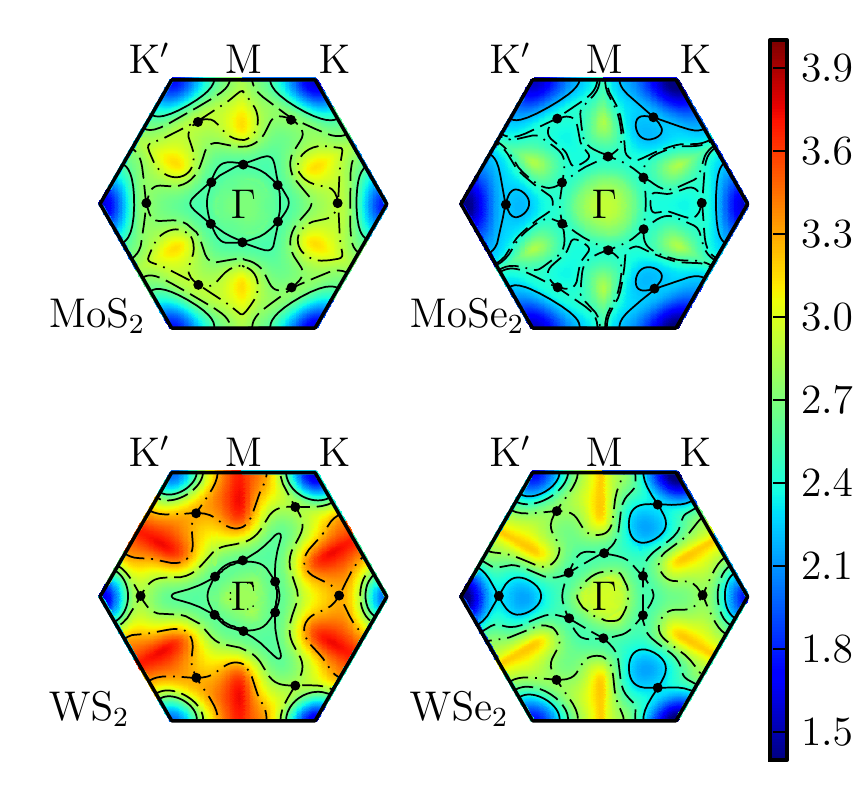}
\caption{(Color online) Color plots of the Wannier-interpolated GGA-PBE band energy difference $\varepsilon_{{\rm c}, \uparrow, \bk}-\varepsilon_{{\rm v}, \uparrow,\bk}$ for the same 2D TMDs as in Figs.~\ref{fig:opt}-\ref{fig:band_diff}. Contour lines of the same quantity are also shown and evaluated at the critical energies at which VHSs occur in the optical conductivity: $E_{1}$ (solid), $E_{2}$ (dashed), and $E_{3}$ (dash-dotted). Filled circles denote the positions of saddle points. \label{fig:contours}}
\end{figure}

In Fig.~\ref{fig:contours} we show color maps of the energy difference 
$\varepsilon_{{\rm c}, \uparrow, \bk}-\varepsilon_{{\rm v}, \uparrow, \bk}$, between the top valence band for spin-$\uparrow$ electrons, $\varepsilon_{{\rm v}, \uparrow, \bk}$, and the bottom conduction band, $\varepsilon_{{\rm c}, \uparrow, \bk}$, for spin-$\uparrow$ electrons. We recall that, by TRS, the results for spin-$\downarrow$ electrons can be simply obtained from the results for spin-$\uparrow$ electrons by sending $\bk$ to $-\bk$, i.e. $\varepsilon_{{\rm c}, \downarrow, \bk}-\varepsilon_{{\rm v}, \downarrow, \bk}  = \varepsilon_{{\rm c}, \uparrow, -\bk}-\varepsilon_{{\rm v}, \uparrow, -\bk}$. We have checked that our numerics respects this important symmetry.

Contour lines of the same quantity are drawn at the critical energies corresponding to the VHSs in the optical conductivity.
For each critical energy we can identify the corresponding saddle point (black filled circles in Fig.~\ref{fig:contours}) and the separatrix line in its neighborhood. 
After a careful inspection of the numerical data, we conclude that, for all 2D TMDs studied in this work, the saddle points are located either along the high symmetry paths 
$\Gamma$-\K and $\Gamma$-\Kp, or {\it very close to the $\Gamma$-\M line}. 
In Fig.~\ref{fig:contours}, the displacement of saddle points from the $\Gamma$-\M line is well visible only in the case of ${\rm MoSe}_2$.
The authors of Refs.~\onlinecite{carvalho_prb_2013,kozawa_naturecommun_2014} claim that in their DFT study 
saddle points occur only along the $\Gamma$-\K/$\Gamma$-\Kp directions. 
We believe that Wannier interpolation is crucial to determine the location of saddle points 
with high accuracy, since this method gives the possibility to increase the $\bk$-space resolution 
much more efficiently than a brute-force DFT method.  

\begin{figure}[t]
\includegraphics{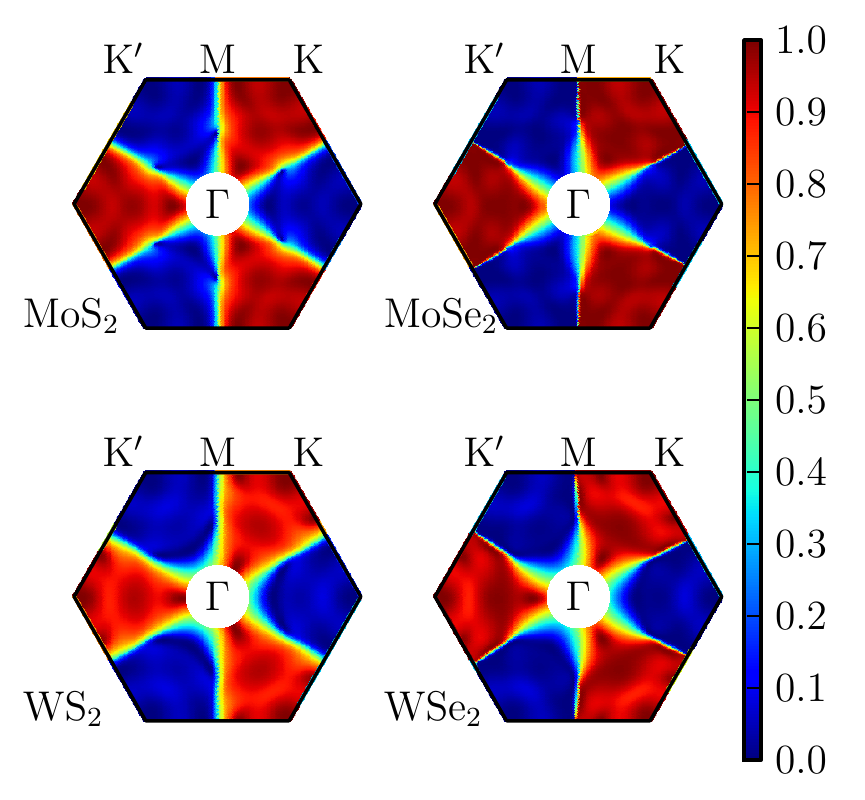}
\caption{(Color online) Color plots of the quantity $\eta^{(+)}_\uparrow(\bk)$ for the same 2D TMDs as in earlier figures. 
Data are not shown in the neighborhood of the $\Gamma$ point due to numerical problems stemming from band crossings occurring at $\Gamma$.\label{fig:polarization}}
\end{figure}

We now note that in Fig.~\ref{fig:opt} certain VHSs in the optical response to left-handed light appear at the same photon energy in both spin channels. Logarithmically large optical responses, however, occur at certain photon energies only in one spin channel. These facts can be explained with the following arguments. We claim that saddle points close to the $\Gamma$-\M line yield identical photo-response in the two spin channels while saddle points along the $\Gamma$-\K/$\Gamma$-\Kp directions yield a substantially spin-polarized optical response. This is due to the matrix elements involved in Eq.~(\ref{eq:spm}). Indeed, we can calculate the spin-resolved amplitude $|\langle {\rm c}, s, \bk|{\hat v}_+ | {\rm v}, s, \bk\rangle|^2$ that enters the optical conductivity in Eq.~(\ref{eq:spm}), i.e.
\begin{equation}\label{eq:eta}
\eta^{(+)}_s(\bk) \equiv \frac{  |\langle {\rm c}, s, \bk|{\hat v}_+ | {\rm v}, s, \bk\rangle|^2}{ |\langle {\rm c}, s, \bk|{\hat v}_+ | {\rm v}, s, \bk\rangle|^2+  |\langle {\rm c}, s, \bk|{\hat v}_- | {\rm v}, s, \bk\rangle|^2}~,
\end{equation}
where the denominator is just a normalization factor. TRS ensures the following relations 
\begin{equation}
\left\{
\begin{array}{l}
\eta^{(-)}_{\downarrow}(\bk) = \eta^{(+)}_{\uparrow}(-\bk)\vspace{0.1 cm}\\
\eta^{(+)}_{\downarrow}(\bk) = 1 - \eta^{(+)}_{\uparrow}(-\bk)
\end{array}
\right.
\end{equation}
between amplitudes for different spin orientations and/or light polarization. The quantity $\eta^{(+)}_\uparrow(\bk)$ is shown in Fig.~\ref{fig:polarization}.

Combining the information in Fig.~\ref{fig:polarization} with that in Fig.~\ref{fig:contours}, we conclude that saddle points along the $\Gamma$-\K ($\Gamma$-\Kp) direction give rise to VHSs only for left-handed (right-
handed) light. 
This means that only one spin component is coupled to light at a given saddle point energy and the corresponding VHS yields a strongly spin-polarized photo-current. 
On the contrary, the matrix elements in Eq.~\eqref{eq:spm} are such that saddle points close to the $\Gamma$-\M line couple 
equally to left- and right-handed light for both spin-$\uparrow$ and spin-$\downarrow$ electrons, so that the corresponding VHSs occur at the same energy.
\section{Summary and conclusions}
\label{sec:end}

In summary, we have presented a fully-relativistic ab-initio density-functional-theory study  of the optical conductivity of 2D group-VIB TMDs.
These calculations have been combined with the use of maximally localized Wannier functions, which offer a computationally inexpensive strategy to reach exceptional ${\bm k}$-space resolution.  

We have focussed on the photo-response of 2D TMDs to circularly-polarized monochromatic light in a wide frequency range, presenting extensive numerical results for monolayer TMDs involving molybdenum and tungsten combined with sulphur and selenium (${\rm MoS}_2$, ${\rm MoSe}_2$, ${\rm WS}_2$, and ${\rm WSe}_2$). We have been able to locate with high accuracy the positions of the points in ${\bm k}$-space that are responsible for van Hove singularities in the optical response. These have been found to occur either along the  high-symmetry directions $\Gamma$-\K and $\Gamma$-\Kp or very close to the $\Gamma$-\M line. Our spin-resolved study provides a route that can be followed experimentally to generate spin-polarized photo-excited carriers by employing circularly polarized light and 2D TMDs. 

In this Article we have neglected electron-electron interactions beyond those described by the LDA or GGA-PBE exchange-correlation energy functional. As we have already mentioned above in Sect.~\ref{sec:bands}, this implies the well-known ``gap problem'', which can be corrected by including electron-electron interactions at the GW level~\cite{shi_prb_2013,louie_prl_2013}. On top of this, one has to keep in mind that low-dimensional systems display large excitonic corrections~\cite{yang_prl_2009,mak_prl_2011,yang_nl_2011}. Shi {\it et al.}~\cite{shi_prb_2013} have calculated a GW energy gap $E_{\rm g}=2.89~{\rm eV}$ and a GW+Bethe Salpeter exciton binding energy $E_{\rm b} = 1.02~{\rm eV}$ for ${\rm MoS}_2$ ($E_{\rm g}=3.02~{\rm eV}$ and $E_{\rm b} = 1.05~{\rm eV}$ for ${\rm WS}_2$). Excitonic effects imply strong optical response at energies close to the gap. Excitonic corrections, however, are not important only at low energies: indeed, they red shift optical absorption peaks related to saddle points. The 
line shape of such VHSs in the optical absorbance is also affected by excitonic corrections, acquiring an asymmetric Fano shape~\cite{kane_pr_1969,yang_prl_2009,mak_prl_2011,yang_nl_2011}. VHSs dressed by excitonic effects have been recently observed in high-quality suspended ${\rm MoS}_2$ devices~\cite{klots_sr_2014}.

To the best of our knowledge, a study of these effects on the spin-resolved optical conductivity tensor ${\hat {\bm \sigma}}(\omega)$ has not yet appeared in the literature and is well beyond the scope of the present work.

\acknowledgements 
M.G. acknowledges partial support by the Max Planck-EPFL Center for Molecular Nanoscience and Technology. 
M.G. and N.M. acknowledge support by a grant from the Swiss National Supercomputing Centre (CSCS) under project ID s337.
F.M.D.P. and M.P. acknowledge support by the E.U. through the Graphene Flagship program (contract no. CNECT-ICT-604391), 
a 2012 SNS Internal Project, and the Italian Ministry of Education, University, and Research (MIUR) through the programs ``FIRB - Futuro in Ricerca 2010" - Project PLASMOGRAPH (Grant No. RBFR10M5BT) and ``Progetti Premiali 2012" - Project ABNANOTECH.

\end{document}